\documentclass[12pt]{article}
\renewcommand{\title}[1]{
\begin{center} \Large \bf #1 \end{center}
}

\renewcommand{\author}[2]{
 \begin{center} #1  \vspace{3mm} \\
  #2 \\
 \end{center}
\addvspace{\baselineskip}
}

\usepackage{amssymb}
\usepackage{amsmath}
\usepackage{multirow,bigdelim}

\usepackage{amsthm}
\newtheorem{thm}{Theorem}[section]
\newtheorem{prop}[thm]{Proposition}
\newtheorem{cor}[thm]{Corollary}
\newtheorem{lem}[thm]{Lemma}
\newtheorem{fact}[thm]{Fact}
\newtheorem{ex}{Example}
\theoremstyle{definition}
\newtheorem{df}{Definition}

\theoremstyle{remark}
\newtheorem*{rem}{Remark}

\theoremstyle{proof}
\newtheorem*{pf}{Proof}

\makeatletter
\@addtoreset{equation}{section}

\makeatother

\date{\today}
\begin{document}

\baselineskip 5mm

\title{
Noncommutative Deformations of Locally Symmetric K\"ahler manifolds
%
}

\author{${}^1$ Kentaro Hara and~ ${}^2$
Akifumi Sako }{
${}^{1,2}$  Department of Mathematics,
Faculty of Science Division II,\\
Tokyo University of Science,
1-3 Kagurazaka, Shinjuku-ku, Tokyo 162-8601, Japan\\
${}^2$  
Fakult\"at f\"ur Physik, Universit\"at Wien\\
Boltzmanngasse 5, A-1090 Wien, Austria}

\abstract{ We derive algebraic recurrence relations to obtain a deformation quantization
with separation of variables for a locally symmetric K\"ahler manifold.
This quantization method is one of the ways to perform a deformation
quantization of K\"ahler manifolds, which is introduced by Karabegov.
From the recurrence relations, concrete expressions of star products for 
one-dimensional local symmetric K\"ahler manifolds  and ${\mathbb C}P^N$ are constructed.
The recurrence relations for a Grassmann manifold $G_{2,2}$ are 
closely studied too.
}

\section{Introduction} 
Deformation quantizations were introduced by Bayen, Flato, Fronsdal, Lichnerowicz 
and Sternheimer \cite{Bayen:1977ha} as a method to quantize spaces.
After \cite{Bayen:1977ha}, several ways of deformation
quantization were proposed \cite{DeW-Lec, Omori, Fedosov, Kontsevich}.
In particular, deformation quantizations of K\"ahler manifolds were
provided in \cite{Moreno86a, Moreno86b, Cahen93, Cahen95}. 
The deformation quantization with separation of variables is one of the methods to construct noncomutative K\"ahler manifolds 
given by Karabegov \cite{Karabegov,Karabegov1996,Karabegov2011}.
In this article, only the deformation quantization with separation of variables
is studied.

A noncomutative product on a quantized manifold is given by a star product
which is defined in a form of a formal power series of deformation parameter $\hbar$.
 The star product is obtained as solutions of an infinite system of differential equations 
. The existence of the solution is proved for a wide class of manifolds,
however explicit expressions of deformation quantizations are constructed only for
few kinds of manifolds, because it is difficult to obtain a concrete solution of the system of differential equations, in general. 

In \cite{Sako:2012ws} explicit expressions of star products of noncomutative ${\mathbb C}P^N$ and ${\mathbb C}H^N$ were provided as the deformation quantization with separation of variables. \footnote{See also the following references.
Star products on the fuzzy
${\mathbb C}P^N$ are investigated in \cite{Balachandran, Kitazawa,
Karabali}. A deformation quantization of the hyperbolic plane was
provided in \cite{Bieliavsky}.}
The reason that they were provided is that ${\mathbb C}P^N$ and ${\mathbb C}H^N$ are locally symmetric spaces.
The symmetry makes our problems be simple ones.
In \cite{Sako:2012ws}, star products on general locally symmetric K\"ahler manifolds are also discussed, but it is not enough to obtain explicit expression of
the star products.\\
\bigskip

The purpose of this article is to derive 
algebraic recurrence relations to make 
concrete expression of star products on locally symmetric K\"ahler manifolds.
The method in this article overcome the problems in \cite{Sako:2012ws}. 
Using the recurrence relations, star products on some 
locally symmetric K\"ahler manifolds, the Riemann surfaces and 
${\mathbb C}P^N$, are obtained.\\

The organization of this article is as follows. In Section \ref{Kahler}, we review the deformation 
quantization with separation of variables proposed by Karabegov.
In Section \ref{locally}, explicit formulas to obtain star products on local symmetric K\"ahler manifolds 
are given explicitly. 
In Section \ref{one}, the explicit expression of star products of one-dimensional locally symmetric K\"ahler manifolds (Riemann surfaces) are constructed.
A two-dimensional case is also discussed.
In Section \ref{grassmann}, we give an explicit formula to obtain star products on a Grassmann manifold, and star products on ${\mathbb C}P^N$
 are obtained by using the formula.
 Finally, we summarize our results and discuss their several perspectives in Section \ref{conclusion}.


\section{Review of the deformation quantization with separation of variables} 
\label{Kahler}

In this section, we review the deformation quantization with
separation of variables 
to construct noncommutative K\"ahler manifolds.

An $N$-dimensional K\"ahler manifold $M$ is described by 
using a K\"ahler potential.
Let $\Phi$ be a K\"ahler potential and $\omega$
be a K\"ahler 2-form:
\begin{eqnarray}
\omega := i g_{k \bar{l}} dz^{k} \wedge d \bar{z}^{l} ,
\ \ 
g_{k \bar{l}} := 
\frac{\partial^2 \Phi}{\partial z^{k} \partial \bar{z}^{l}} . \label{2_1}
\end{eqnarray}
where $ z^i,\bar{z}^i \ (i=1,2,\dots , N)$ are complex local coordinates. \\
In this article, we use the Einstein summation convention over repeated
indices. 
The $g^{\bar{k} l}$ is the inverse of the K\"ahler metric tensor $g_{k \bar{l}}$. That means 
$ g^{\bar{k} l}  g_{l \bar{m}} = \delta_{\bar{k} \bar{m} } $.
In the following, we use 
\begin{align}
\partial_k = \frac{\partial}{\partial z^{k}} , ~
\partial_{\bar{k}} = \frac{\partial}{\partial \bar{z}^{k}}.
\end{align}

Deformation quantization is defined as follows.
\begin{df}[Deformation quantization]
Deformation quantization of Poisson manifolds is defined as follows.
$\cal F$ is defined as a set of formal power series:
$
{\cal F} := \left\{  f \ \Big| \ 
f = \sum_k f_k \hbar^k, ~f_k \in C^\infty \left(M \right)
\right\} .
$
A star product is defined as 
\begin{eqnarray}
f * g = \sum_k C_k (f,g) \hbar^k
\end{eqnarray}
such that the product satisfies the following conditions.
\begin{enumerate}
\item $\left({\cal F},+, *\right)$ is a (noncommutative) algebra.
\item $C_k\left(\cdot ,\cdot  \right)$ is a bidifferential operator.
\item $C_0$ and $C_1$ are defined as 
\begin{eqnarray}
&& C_0 (f,g) = f g,  \\
&&C_1(f,g)-C_1(g,f) = \{ f, g \}, \label{weakdeformation}
\end{eqnarray}
where $\{ f, g \}$ is the Poisson bracket.
\item $ f * 1 = 1* f = f$.
\end{enumerate}
\end{df}

Karabegov introduced a method to obtain a deformation quantization of a K\"ahler manifold in \cite{Karabegov1996}.
His deformation quantization is called deformation quantizations with separation of variables
\begin{df}[A star product with separation of variables]
$*$ is called a star product with separation of variables on a K\"ahler manifold when 
\begin{eqnarray}
a * f = a f 
\end{eqnarray}
for an arbitrary holomorphic function $a$ and
\begin{eqnarray} 
f * b = f b
\end{eqnarray}
for an arbitrary anti-holomorphic function $b$.
\end{df}

We use 
$$
D^{\bar{l}} = g^{\bar{l} k } \partial_k
$$
and introduce 
$$
{\cal S} := 
\Big\{ 
A~ |~ A=\sum_{\alpha} a_{\alpha} D^{\alpha} ,\ \ 
a_{\alpha} \in C^{\infty} \left(M \right)
\Big\} ,
$$
where ${\alpha}$ is a multi-index 
$\alpha = (\alpha_1 , \alpha_2 , \dots , \alpha_n)$.
In this article, we also use the Einstein summation convention over repeated multi-indices 
and $a_{\alpha} D^{\alpha}:=\sum_{\alpha} a_{\alpha} D^{\alpha}$. 

There are some useful formulae.
$D^{\bar{l}}$ satisfies the following equations.
\begin{align}
[ D^{\bar{l}} , D^{\bar{m}} ] = 0 \ \ ,\ \  
[ D^{\bar{l}} , \partial_{\bar{m}} \Phi ] 
&= {\delta^{\bar{l}}}_{\bar{m}} ,\ \ \forall l,m , \label{lem1-1}
\end{align}
where $[A, ~B] = AB - BA$ .
Using them, one can construct
a star product as a differential operator $L_f$ such that
$f*g=L_f g$.
\begin{thm} \label{theo1}
[Karabegov \cite{Karabegov1996}]. For an arbitrary K\"ahler form $\omega$, there exist a star product with separation of
variables $*$ and it is constructed as follows.  Let $f$ be an element
of ${\cal F}$ and $A_n \in {\cal S}$ be a differential operator whose
coefficients depend on $f$ i.e.
\begin{eqnarray}
A_n = a_{n, \alpha}(f) D^{\alpha} , \ 
D^{\alpha}= \prod_{i=1}^n (D^{\bar{i}})^{\alpha_i} , \ 
(D^{\bar{i}}) =g^{\bar{i} l} \partial_l ,
\end{eqnarray}
where 
$\alpha$ is an multi-index $\alpha = (\alpha_1 , \alpha_2 , \dots , \alpha_n)$.
Then,
\begin{eqnarray}
L_f = \sum_{n=0}^{\infty} \hbar^n A_n 
\end{eqnarray}
is uniquely determined such that
it satisfies the following conditions.
\begin{enumerate}
\item For $\displaystyle R_{\partial_{\bar{l}} \Phi} = 
\partial_{\bar{l}}\Phi + \hbar \partial_{\bar{l}}$, 
\begin{eqnarray}
\left[  L_f , R_{\partial_{\bar{l}} \Phi} \right]=0 \ .
\label{A_f_R_Phi}
\end{eqnarray}
\item 
\begin{eqnarray}
L_f 1 &=& f*1=f . \label{A_f_2_1}
\end{eqnarray}
\end{enumerate}
Then the star products are given by
\begin{eqnarray}
L_f g &:=& f * g , \label{A_f_2_2}
\end{eqnarray}
and the star products satisfy the associativity;
\begin{eqnarray}
L_h ( L_g f ) &=& h * (g * f)
= (h*g)*f = L_{L_h g} f . \label{A_f_2_3}
\end{eqnarray}

\end{thm}
Recall that each two of $D^{\bar{i}}$ 
commute each other, so if a multi index $\alpha$ is fixed
then the $A_n$ is uniquely determined.
(\ref{A_f_2_1})-(\ref{A_f_2_3})
imply that $L_f g =f * g$
gives deformation quantization. 



\begin{df} A map from differential operators to formal polynomials is defined as
\[
\sigma \left(A;\xi\right):=\sum _\alpha a_\alpha \xi^\alpha,
\]
where
\[
A=\sum _\alpha a_\alpha D^\alpha.
\]
\end{df}
This map is called ``twisted symbol''. 
It becomes easier to calculate commutators by using the following theorem.

\begin{prop}[Karabegov \cite{Karabegov1996}]
 Let $a(\xi)$ be a twisted symbol of an operator
 $A$. Then the twisted symbol of the operator
 $[A, \partial_{\bar{i}}\Phi] $ is equal to $\partial
a/\partial\xi^{\bar{i}}$; \label{symbol}
\[
\sigma \left([A, \partial_{\bar{i}}\Phi] \right)=\frac{\partial }{\partial \xi^{\bar{i}}}\sigma \left(A\right) .
\]
\end{prop}
This proposition follows from 
From {\rm(\ref{lem1-1})}, i.e.
\[
\sigma\left([D^{\bar{l}}, \partial_{\bar{i}}\Phi] \right)
=\delta^{\bar{l}}_{\: \bar{i}} . 
\]

\section{Deformation quantization with separation of variables for a locally symmetric K\"ahler manifold}
\label{locally}
In this section, explicit formulas to obtain star products on local symmetric K\"ahler manifolds are constructed. 
A method of Karabegov in Section \ref{Kahler} is used for the constructing.\\

At first we list notations used in this article.
Let $M$ be a $N$-dimensional K\"ahler manifold, $\displaystyle \partial _i:=\frac{\partial}{\partial z_i},
\partial _{\bar{i}}:=\frac{\partial}{\partial \bar{z}_i} ~ (i= 1,\dots , N) $ be tangent vector fields on a coordinate chart $U\subset M$ 
with its local coordinates $\left(z^1,\cdots ,z^N,\bar{z}^1,\cdots ,\bar{z}^N \right),dz^i, d\bar{z}^i$ be cotangent vector fields on $U$ and 
${Y^{\mu_1\cdots \mu_k\bar{\mu}_1\cdots \bar{\mu}_{m}}}_{\nu_1\cdots \nu_l\bar{\nu}_1\cdots \bar{\nu}_n} 
{\partial_{\mu_1}} \otimes \cdots \otimes {\partial_{\mu_k}} \otimes{\partial_{\bar{\mu}_1}} \otimes  \cdots \otimes {\partial_{\bar{\mu}_m}}
 \otimes dz^{\nu_1} \otimes \cdots \otimes dz^{\nu_l}\otimes d\bar{z}^{\bar{\nu}_1} \otimes \cdots \otimes d\bar{z}^{\bar{\nu}_n}
 \\ \in \Gamma\left[\left( T^{1,0}M \right)^{\otimes k}\otimes  \left( T^{0,1}M \right)^{\otimes m}\otimes 
 \left\{ \left(T^{1,0}M \right)^* \right\}^{\otimes l}\otimes \left\{ \left(T^{0,1}M \right)^* \right\}^{\otimes n} \right] $
be a $(\left(k,m \right),\left(l,n \right))$-tensor field.
The classical style of covariant derivative $\nabla_i:=\nabla_{\partial_i}$ acts on coefficients of tensor fields as
\begin{align*}
\lefteqn{\nabla_i{Y^{\mu_1\cdots \mu_k\bar{\mu}_1\cdots \bar{\mu}_{m}}}_{\nu_1\cdots \nu_l\bar{\nu}_1\cdots \bar{\nu}_n}} \\
 &=\partial_i{Y^{\mu_1\cdots \mu_k\bar{\mu}_1\cdots \bar{\mu}_{m}}}_{\nu_1\cdots \nu_l\bar{\nu}_1\cdots \bar{\nu}_n} \\
&+\sum _{q=1}^k\Gamma^{\mu_q}_{i\rho_q}{Y^{\mu_1\mu_2\cdots \rho_q\cdots \mu_k\bar{\mu}_1\cdots \bar{\mu}_{m}}}_{\nu_1\cdots \nu_l\bar{\nu}_1\cdots \bar{\nu}_n}
+\sum _{q=1}^m\Gamma^{\bar{\mu}_q}_{i\bar{\rho}_q}{Y^{\mu_1\cdots \mu_k\bar{\mu}_1\bar{\mu}_2\cdots \bar{\rho_q}\cdots \bar{\mu}_{m}}}_{\nu_1\cdots \nu_l\bar{\nu}_1\cdots \bar{\nu}_n}\\
 &\qquad -\sum _{q=1}^l\Gamma^{\sigma_q}_{i\nu_q}{Y^{\mu_1\cdots \mu_k\bar{\mu}_1\cdots \bar{\mu}_{m}}}_{\nu_1\nu_2\cdots \sigma_q \cdots \nu_l\bar{\nu}_1\cdots \bar{\nu}_n} 
 -\sum _{q=1}^n\Gamma^{\bar{\sigma}_q}_{i\bar{\nu}_q}{Y^{\mu_1\cdots \mu_k\bar{\mu}_1\cdots \bar{\mu}_{m}}}_{\nu_1\cdots \nu_l\bar{\nu}_1\bar{\nu}_2\cdots \bar{\sigma}_q\cdots \bar{\nu}_n}
\end{align*}
where $\Gamma^{i}_{jk}$ is the Christoffel symbol.

The Riemannian curvature of a Hermitian manifold $M$ is defined as
\[
{R_{i\bar{j}k}}^l =\partial_i\Gamma^l_{\bar{j}k}  - \partial_{\bar{j}}\Gamma^l_{ik}  +\Gamma^n_{\bar{j}k} \Gamma^l_{in} -\Gamma^n_{ik}\Gamma^l_{\bar{j}n} 
.\]
For Hermitian manifolds, the Christoffel symbols are given as 
\[
\Gamma^l_{jk}=g^{l\bar{q}}\frac{\partial g_{j\bar{q}} }{\partial z^k}.
\]

The Riemannian  curvature of a Hermitian manifold $M$ is obtained as
\[
R_{i\bar{j} k \bar{l}}= -\frac{\partial^2 g_{\bar{j}i}}{\partial z^k\partial \bar{z}^l}
 +g^{p\bar{q}}\frac{\partial g_{i\bar{q}} }{\partial z^k}\frac{\partial g_{\bar{j}p}}{\partial \bar{z}^l}.
\]
On a K\"ahler manifold, its metric is described by using K\"ahler potential $\Phi$ as (\ref{2_1}).
Then its Riemannian  curvature is given by
\begin{equation}\label{symr}
R_{i\bar{j} k \bar{l}}=-\frac{\partial^4  \Phi}{\partial z^i\partial \bar{z}^j\partial z^k\partial \bar{z}^l}
 +g^{p\bar{q}}\frac{\partial^3  \Phi }{\partial z^i\partial \bar{z}^q\partial z^k}
 \frac{\partial^3  \Phi}{\partial z^p\partial \bar{z}^j\partial \bar{z}^l}.
\end{equation}

(See \cite{Kobayashi_Nomizu} P157.)

Operators $D^{\vec{\alpha_n}}$ and $D^{\vec{\beta_n^*}}$ are defined 
by using $D^k=g^{k\bar{m}} \partial _{\bar{m}}$ and
$D^{\bar{j}}=g^{\bar{j}l}\partial _l $ as

\[
D^{\vec{\alpha_n}}:=D^{\alpha^n_1}D^{\alpha^n_2}\cdots D^{\alpha^n_N}
\qquad ,\qquad \: D^{\vec{\beta_n}}:=D^{\beta_1}D^{\beta_2}\cdots D^{\beta_N}
\]
where
\[
D^{\alpha_k}:=\left(D^k \right)^{\alpha_k}\qquad ,\qquad  D^{\beta_j}:=\left(D^{\bar{j}} \right)^{\beta_j},
\]
and $\vec{\alpha_n}$ and $\vec{\beta^*_n}$ are
$N$-dimensional vectors whose summation of their all elements are set to be $n$;
\begin{align*}
\vec{\alpha}_n
\in \left\{\left(\gamma^n_1,\gamma^n_2,\cdots ,\gamma^n_N \right)
\in\mathbb{Z}^N ~
\Big| ~ \sum _{k=1}^N\gamma^n_k=n  \right\} , \ \
 \vec{\beta_n^*}
\in \left\{\left(\gamma^n_1,\gamma^n_2,\cdots ,\gamma^n_N \right)^*
\in\mathbb{Z}^N ~
\Big| ~ \sum _{k=1}^N\gamma^n_k=n  \right\} 
\end{align*}
i.e.
\begin{align*}
&\vec{\alpha_n}:=\left(\alpha^n_1,\alpha^n_2,\cdots ,\alpha^n_N \right),
\ \ |\vec{\alpha}_n |:= \sum _{k=1}^N\alpha^n_k=n  \\
&\vec{\beta_n^*}:=\left(\beta^n_1,\beta^n_2,\cdots ,\beta^n_N \right)^*,\ \ 
|\vec{\beta^*_n}|:= \sum _{k=1}^N\beta^n_k=n .
\end{align*}
For $\vec{\alpha_n}\notin \mathbb{Z}_{\ge 0}^N$ 
we define $D^{\vec{\alpha_n}}:=0$.

For example, 
$D^{\left(1,2,3 \right)}=D^1\left(D^2 \right)^2\left(D^3 \right)^3$
,~ $D^{{\left(2,4,0 \right)}^*}=\left(D^{\bar{1}} \right)^2\left(D^{\bar{2}} \right)^4$
and $D^{\left(5,-2,3 \right)}=0$ for a $3$-dimensional manifolds case
with $n=6$.

$\vec{e_i}$ is used as a $N$-dimensional vector
\begin{align}
\vec{e_i} = ( \delta_{1i} , \delta_{2i} , \cdots , \delta_{Ni} ).
\end{align}

From here to the end of this section, we make up recurrence relations to construct
 explicit expressions of star products on locally symmetric K\"ahler manifolds.

A Riemannian(K\"ahler) manifold $\left(M,g \right)$ is called a locally symmetric Riemannian(K\"ahler) manifold when $\nabla_m {R_{ijk}}^l=0 ~
(\forall i,j,k,l,m )$.
Only locally symmetric K\"ahler manifolds are studied in this article.\\

We assume that a star product with separation of variables for smooth functions $f$ and $g$ on a locally symmetric K\"ahler manifold $M$ has a form
\begin{equation}\label{lfg}
L_fg=f * g =\sum_{n=0}^\infty \sum_{\substack{\vec{\alpha_n} \vec{\beta_n^*}  \\ }}
T^n_{\vec{\alpha_n}\vec{\beta_n^*}} \left(D^{\vec{\alpha_n}} f\right)  \left(D^{\vec{\beta_n^*}}g\right) ,
\end{equation}
where $T^n_{\vec{\alpha_n}\vec{\beta_n^*}}$ are covariantly constants. 
If $\vec{\alpha_n}\notin \mathbb{Z}_{\ge 0}^N$ or $\vec{\beta_n}\notin \mathbb{Z}_{\ge 0}^N$ then we define $T^n_{\vec{\alpha_n}\vec{\beta_n^*}} :=0$.
$\displaystyle\sum_{\substack{\vec{\alpha_n} \vec{\beta_n^*}  \\ }}$ is defined by the summation over all 
$\vec{\alpha_n^*}$ and $\vec{\beta_n^*}$ satisfying $\left|\vec{\alpha_n^*} \right|=\: \left|\vec{\beta_n^*} \right|=n$.
In brief,
\[
n=\left|\vec{\alpha_n^*} \right|:=\sum_{i=1}^N\alpha^n_i ,\ \  
n=\: \left|\vec{\beta_n^*} \right|:=\sum_{i=1}^N\beta^n_i 
\: ,\ \   
\sum_{\substack{\vec{\alpha_n} \vec{\beta_n^*}  \\ }}:=\sum_{\substack{\left|\vec{\alpha_n} \right|=\left|\vec{\beta_n^*} \right|=n \\ }}.
\]
\begin{prop}\label{t1}
For the star product  on a locally symmetric K\"ahler manifold $M$ as (\ref{lfg}), $T^0_{\vec{\alpha_0}\vec{\beta_0^*}} $ and 
$T^1_{\vec{e_i},\vec{e_j}}$ are given as
\[
T^0_{\vec{\alpha_0}\vec{\beta_0^*}} = 1 , \ \ 
T^1_{\vec{e_i},\vec{e_j}} =\hbar g_{i\bar{j}}.
\]
\end{prop}
\begin{pf} From (\ref{lfg}), the star product for smooth functions $f$ and $g$ on $M$ is given as 
\[
L_fg=T^0_{\vec{\alpha_0}\vec{\beta_0^*}} fg
+ \sum_{\substack{\vec{\alpha_1} \vec{\beta_1^*}}} T^1_{\vec{\alpha_1}\vec{\beta_1^*}} \left(D^{\vec{\alpha_1}} f\right) \left(D^{\vec{\beta_1^*}}g\right) 
+\sum_{n=2}^\infty \sum_{\substack{\vec{\alpha_n} \vec{\beta_n^*}}} T^n_{\vec{\alpha_n}\vec{\beta_n^*}} \left(D^{\vec{\alpha_n}} f\right) \left(D^{\vec{\beta_n^*}}g\right) .
\]
$T^0_{\vec{\alpha_0}\vec{\beta_0^*}} =1$ is trivial.
$C_1(f,g)$ is expressed as
\[
C_1(f,g)=\sum_{\substack{\vec{\alpha_1} \vec{\beta_1^*}}}T^1_{\vec{\alpha_1}\vec{\beta_1^*}} \left(D^{\vec{\alpha_1}} f\right) \left(D^{\vec{\beta_1^*}}g\right).
\]
By the definition of the deformation quantization (\ref{weakdeformation}) 
the first term is related to the Poisson bracket:
\begin{align*}
&\hbar \sum_{i,j=1}^ng^{i\bar{j}}\left(\frac{\partial f}{\partial z^i}\frac{\partial g}{\partial \bar{z}^j}
-\frac{\partial g}{\partial \bar{z}^j}\frac{\partial f}{\partial z^i}  \right)  \\
 &=
\sum_{\substack{\vec{\alpha_1} \vec{\beta_1^*}}}
(T^1_{\vec{\alpha_1}\vec{\beta_1^*}} \left(g^{\alpha_1\bar{m}}\partial _{\bar{m}} f\right) \left(g^{\bar{\beta_1}l}\partial _lg\right) 
-T^1_{\vec{\alpha_1}\vec{\beta_1^*}} \left(g^{\alpha_1\bar{m}}\partial _{\bar{m}}  g\right) \left(g^{\bar{\beta_1}l}\partial _lf\right) ).
\end{align*}
Then $T^1_{\vec{e_i},\vec{e_j}} =\hbar g_{i\bar{j}}$ is shown.
\qed\end{pf}

The purpose of remained part of this section is to replace the recurrence relations as differential equations by those of algebraic equations.
We need to calculate $\left[L_f, \partial_{\bar{i}}\Phi\cdot \right]$ and $\left[L_f,\partial_{\bar{i}} \right]$ in (\ref{A_f_R_Phi}).
\begin{prop}
Let $f$ and $g$ be smooth functions on a locally symmetric K\"ahler manifold $M$and $L_f$ be a left star product 
by $f$ given as (\ref{lfg}). Then\label{phi}
\begin{align*}
\lefteqn{\sigma \left([L_f, \partial_{\bar{i}}\Phi  ] \right)=\frac{\partial \sigma \left(L_f\right)}{\partial \xi^{\bar{i}}}} \\
 &= \left \{
\begin{array}{l}
\sum_{n=0}^\infty \sum_{\vec{\alpha_n} \vec{\beta_n^*}}\beta_i^n T^n_{\vec{\alpha_n}\vec{\beta_n^*}} \left(D^{\vec{\alpha_n}} f\right) 
 \left({\xi^{\bar{1}}}^{\beta_1^n}\cdots{\xi^{\bar{i}}}^{\beta_i^n-1}\cdots {\xi^{\bar{N}}}^{\beta_N^n} \right)\qquad \left(\beta_i\neq 0 \right)\\
0\qquad \left(\beta_i^n=0 \right)
\end{array}
\right. ,
\end{align*}
or equivalently,
\begin{align}
\left[L_f, \partial_{\bar{i}}\Phi\right]g=
\left\{
\begin{array}{ll}
\sum_{n=0}^\infty \sum_{\substack{\vec{\alpha_n} \vec{\beta_n^*}}} \beta_i^n T^n_{\vec{\alpha_n}\vec{\beta_n^*}} \left(D^{\vec{\alpha_n}} f\right) 
 \left(D^{\vec{\beta_n^*}-\vec{e_i}}g \right)  \left(\beta_i\neq 0 \right)\\
0\qquad \left(\beta_i^n=0 \right)
\end{array}
\right. . \label{tuika1}
\end{align}
\end{prop}

\begin{pf} By Proposition\, \ref{symbol},
\[
\sigma \left([L_f, \partial_{\bar{i}}\Phi\cdot ] \right)=\frac{\partial \sigma \left(L_f\right)}{\partial \xi^{\bar{i}}}
= \frac{\partial}{\partial \xi^{\bar{i}}} \sum_{n=0}^\infty 
\sum_{\substack{\vec{\alpha_n} \vec{\beta_n^*}}} 
T^n_{\vec{\alpha_n}\vec{\beta_n^*}} 
 \left(D^{\vec{\alpha_n}} f\right) \left(\xi^{\vec{\beta_n^*}}\right) .
\]
$\xi^{\vec{\beta_n^*}}$ is explicitly written by 
$\displaystyle \xi^{\vec{\beta_n^*}}= {\xi^{\bar{1}}}^{\beta_1^n}{\xi^{\bar{2}}}^{\beta_2^n}\cdots {\xi^{\bar{N}}}^{\beta_N^n} $, then
\begin{align*}
&\lefteqn{\sigma \left([L_f, \partial_{\bar{i}}\Phi\cdot ] \right)} \\
 &= \left \{
\begin{array}{l}
\sum_{n=0}^\infty\sum_{\vec{\alpha_n} \vec{\beta_n^*}} \beta_i^n T^n_{\vec{\alpha_n}\vec{\beta_n^*}} \left(D^{\vec{\alpha_n}} f\right) 
 \left({\xi^{\bar{1}}}^{\beta_1^n}\cdots{\xi^{\bar{i}}}^{\beta_i^n-1}\cdots {\xi^{\bar{N}}}^{\beta_N^n} \right)\qquad \left(\beta_i\neq 0 \right)\\
0\qquad \left(\beta_i=0 \right)
\end{array}
\right. .
\end{align*}
\qed\end{pf}

The following formulas are given in \cite{Sako:2013noa}.
\begin{fact}\label{nabla}
For smooth functions $f$ and $g$ on a locally symmetric K\"ahler manifold,
the following formulas are given.
\begin{align*}
\lefteqn{ \nabla_{\bar{j}_1} \cdots \nabla_{\bar{j}_n} f=  g_{l_1 \bar{j}_1} \cdots g_{l_n \bar{j}_n} D^{l_1} \cdots D^{l_n} f} \\
 &\nabla_{k_1} \cdots \nabla_{k_n} g= g_{\bar{m}_1 k_1} \cdots g_{\bar{m}_n k_n} D^{\bar{m}_1} \cdots D^{\bar{m}_n} g \\
 & D^{l_1} \cdots D^{l_n} f = g^{l_1 \bar{j}_1} \cdots g^{l_n \bar{j}_n} \nabla_{\bar{j}_1} \cdots \nabla_{\bar{j}_n}f\\
 & D^{\bar{m}_1} \cdots D^{\bar{m}_n} g=g^{\bar{m}_1 k_1} \cdots g^{\bar{m}_n k_n} \nabla_{k_1} \cdots \nabla_{k_n} g .
\end{align*}
\end{fact}

Fact \ref{nabla} derives the following lemma.
\begin{lem}Let $f$ and $g$ be smooth functions on a locally symmetric K\"ahler manifold $M$.
Let $L_f$ be a left star product by $f$ given as (\ref{lfg}).
Then,
\label{partial}
\begin{align*}
\lefteqn{[L_f,\hbar \partial_{\bar{i}}]g } \\
  &=\hbar\sum_{n=0}^\infty \sum_{\vec{\alpha_n} \vec{\beta_n^*}}\sum _{k=1}^N\sum_{\vec{\alpha_n} \vec{\beta_n^*}}
  \frac{\beta^n_k\left(\beta^n_k-1 \right)}{2}{{R_{\bar{\rho}}}^{\bar{k}\bar{k}}}_{\bar{i}}
  \; T^n_{\vec{\alpha_n}\vec{\beta_n^*}}\left(D^{\vec{\alpha_n}}f \right)\left(  D^{\vec{\beta_n^*}+\vec{e_\rho}-\vec{e_k}}g \right)\\
  &\quad +\hbar\sum_{n=0}^\infty \sum _{k=1}^{N-1}\sum _{l=1}^{N-k}\sum_{\vec{\alpha_n} \vec{\beta_n^*}}
  \beta^n_k\beta^n_{k+l} {{R_{\bar{\rho}}}^{\overline{k+l}\bar{k}}}_{\bar{i}}\;
  T^n_{\vec{\alpha_n}\vec{\beta_n^*}} \left( D^{\vec{\alpha_n}}f \right) \left( D^{\vec{\beta_n^*}+\vec{e_\rho}-\vec{e_k}}g\right)\\
&\quad -\hbar\sum_{n=1}^\infty \sum_{\vec{\alpha_{n-1}} \vec{\beta_{n-1}^*}} 
\sum _{d=1}^N g_{\bar{i}d}T^{n-1}_{\vec{\alpha_{n-1}}\vec{\beta_{n-1}^*}}
\left( D^{\vec{\alpha_{n-1}}+\vec{e_d}}f \right)\left( D^{\vec{\beta_{n-1}^*}}g  \right) .
\end{align*}
\end{lem}
\begin{pf}
We can calculate $[L_f,\hbar \partial_{\bar{i}}]g$ straightforwardly.
\begin{align}
[L_f,\hbar \partial_{\bar{i}}]g=&\hbar [L_f,\nabla_{\bar{i}}]g
\notag \\
=&
\lefteqn{\hbar\sum_{n=0}^\infty \sum_{\vec{\alpha_n} \vec{\beta_n^*}}
T^n_{\vec{\alpha_n}\vec{\beta_n^*}} \left\{\left( D^1\right)^{\alpha_1^n}\cdots \left(D^N \right)^{\alpha_N^n}f \right\} \left[
\left(D^{\bar{1}} \right)^{\beta_1^n}\cdots \left(D^{\bar{N}} \right)^{\beta_N^n},\nabla_{\bar{i}}  \right]g} \nonumber \\  
&\qquad -\hbar\sum_{n=0}^\infty \sum_{\vec{\alpha_n} \vec{\beta_n^*}}
T^n_{\vec{\alpha_n}\vec{\beta_n^*}}\left\{\nabla_{\bar{i}}\left( D^1\right)^{\alpha_1^n}\cdots \left(D^N \right)^{\alpha_N^n}f \right\}
\left\{\left(D^{\bar{1}} \right)^{\beta_1^n}\cdots \left(D^{\bar{N}} \right)^{\beta_N^n}g \right\}.\label{lfd1}
\end{align}

From Fact \ref{nabla}, the second term of (\ref{lfd1}) becomes
\begin{align}
\lefteqn{\hbar\sum_{n=0}^\infty \sum_{\vec{\alpha_n} \vec{\beta_n^*}}T^n_{\vec{\alpha_n}\vec{\beta_n^*}}
\left\{\nabla_{\bar{i}}\left( D^1\right)^{\alpha^n_1}\cdots \left(D^N \right)^{\alpha^n_N}f \right\}
 \left\{\left(D^{\bar{1}} \right)^{\beta^n_1}\cdots \left(D^{\bar{N}} \right)^{\beta^n_N}g \right\}} \nonumber  \\
 &=\hbar\sum_{n=1}^\infty  \sum _{d=1}^N\sum_{\vec{\alpha_{n-1}} \vec{\beta_{n-1}^*}}g_{\bar{i}d}T^{n-1}_{\vec{\alpha_{n-1}}\vec{\beta_{n-1}^*}}
 \left( D^{\vec{\alpha_{n-1}}+\vec{e_d}}f \right)\left( D^{\vec{\beta_{n-1}^*}}g  \right).  \label{ddf}
\end{align}

To calculate the first term of (\ref{lfd1}) we calculate $\left[D^{\vec{\beta_n^*}} ,\nabla_{\bar{i}}  \right]g :$
\begin{align}
\lefteqn{\left[\left(D^{\bar{1}} \right)^{\beta^n_1}\cdots \left(D^{\bar{N}} \right)^{\beta^n_N},\nabla_{\bar{i}}  \right]g } \nonumber  \\
 &= \left\{\left[\left(D^{\bar{1}} \right)^{\beta^n_1},\nabla_{\bar{i}}  \right] \left\{\left(D^{\bar{2}} \right)^{\beta^n_2}\cdots \left(D^{\bar{N}} \right)^{\beta^n_N} \right\}
+\cdots +\left\{\left(D^{\bar{1}} \right)^{\beta^n_1}\cdots \left(D^{\overline{N-1}} \right)^{\beta^n_{N-1}} \right\}
\left[\left(D^{\bar{N}} \right)^{\beta^n_N},\nabla_{\bar{i}}  \right] \right\}g.  \nonumber 
 \end{align}

For these terms, we evaluate
\begin{equation}\label{dddg}
\left[\left(D^{\bar{a}} \right)^{\beta^n_a},\nabla_{\bar{i}}  \right] 
\left(D^{\overline{a+1}} \right)^{\beta^n_{a+1}}\cdots \left(D^{\bar{N}} \right)^{\beta^n_N}g
=\sum _{m=1}^{\beta^n_a}\left(D^{\bar{a}} \right)^{m-1}\left[D^{\bar{a}},\nabla_{\bar{i}}  \right] \left(D^{\bar{a}} \right)^{\beta^n_a-m}
 \left(D^{\overline{a+1}} \right)^{\beta^n_{a+1}}\cdots \left(D^{\bar{N}} \right)^{\beta^n_N}g
\end{equation}
by cases, Case1 $\beta^n_a =1$, Ace2 $\beta^n_a >1$ and $\: \sum_{k=a+1}^N \beta^n_k>0$, and
Cave3 $\beta^n_a >1$ and $\: \sum_{k=a+1}^N \beta^n_k=0$.\\

Cache1.  If $\beta^n_a =1$ the last line of (\ref{dddg}) is written as
\begin{align}
\lefteqn{\left[\left(D^{\bar{a}} \right)^{\beta^n_a},\nabla_{\bar{i}}  \right] 
\left(D^{\overline{a+1}} \right)^{\beta^n_{a+1}}\cdots \left(D^{\bar{N}} \right)^{\beta^n_N}g}  \nonumber \\
 &=\sum_{j=a+1}^N \sum_{n_j=1}^{\beta^n_j} {{R_{\bar{i}}}^{\bar{a}\bar{j}}}_{\bar{c}}\; D^{\bar{c}}\left(D^{\overline{a+1}} \right)^{\beta^n_{a+1}}
\cdots \left(D^{\bar{j}} \right)^{\beta^n_j-1}\cdots \left(D^{\bar{N}} \right)^{\beta^n_N} g  \nonumber \\ 
 &=\sum_{j=a+1}^N \beta^n_j \beta^n_a{{R_{\bar{i}}}^{\bar{a}\bar{j}}}_{\bar{c}}\; D^{\bar{c}}
\left(D^{\bar{a}} \right)^{\beta^n_a-1} \left(D^{\overline{a+1}} \right)^{\beta^n_{a+1}}
\cdots \left(D^{\bar{j}} \right)^{\beta^n_j-1}\cdots \left(D^{\bar{N}} \right)^{\beta^n_N} g. \label{dr}
\end{align}
Recall that $(D^{\bar{j}})^n =0 $ for negative $n$ by definition.\\

Carce2. If $\beta^n_a >1$ and $\: \sum_{k=a+1}^N \beta^n_k>0$, by using Fact \ref{nabla}, we obtain (\ref{dddg}) as
\begin{align}\label{S}
\lefteqn{\sum _{m=1}^{\beta^n_a} \left(D^{\bar{a}} \right)^{m-1}\left[D^{\bar{a}},\nabla_{\bar{i}}  \right]  
\left(D^{\bar{a}} \right)^{\beta^n_a-m}\left(D^{\overline{a+1}} \right)^{\beta^n_{a+1}}\cdots \left(D^{\bar{N}} \right)^{\beta^n_N}g} \\
=&\sum _{m=1}^{\beta^n_a} \left(D^{\bar{a}} \right)^{m-1}g^{\bar{a}b}g^{\bar{a}k_{a,1}} \cdots g^{\bar{a}k_{a,\beta^n_a-m} } \left[\nabla_b,\nabla_{\bar{i}}  \right] 
 \nabla_{k_{a,1}} \cdots \nabla_{k_{a,\beta^n_a-m}}
\left(D^{\overline{a+1}} \right)^{\beta^n_{a+1}}\cdots \left(D^{\bar{N}} \right)^{\beta^n_N}g \notag \\
=&\lefteqn{\sum _{m=1}^{\beta^n_a}\sum_{n_a=1}^{\beta^n_a-m} \left(D^{\bar{a}} \right)^{m-1} {{R_{\bar{i}}}^{\bar{a}\bar{a}}}_{\bar{c}}\;  D^{\bar{c}}
\left(D^{\bar{a}} \right)^{\beta^n_a-m} \left(D^{\overline{a+1}} \right)^{\beta^n_{a+1}}\cdots \left(D^{\bar{N}} \right)^{\beta^n_N}g } \nonumber \\
& + \sum _{m=1}^{\beta^n_a}\sum_{j=a+1}^N \sum_{\vec{\alpha_n} \vec{\beta_n^*}}\beta^n_j{{R_{\bar{i}}}^{\bar{a}\bar{j}}}_{\bar{c}}\; D^{\bar{c}}
\left(D^{\bar{a}} \right)^m \left(D^{\overline{a+1}} \right)^{\beta^n_{a+1}}
\cdots \left(D^{\bar{j}} \right)^{\beta^n_j-1}\cdots \left(D^{\bar{N}} \right)^{\beta^n_N} g .
\end{align}
Here, we used 
\begin{equation}\label{R}
\left[\nabla_i,\nabla_j \right]\nabla_{k_1} \cdots \nabla_{k_m}f
=-\sum_{n=1}^m {R_{ijk_n}}^l\; \nabla_{k_1} \cdots \nabla_{k_{n-1}} \nabla_l \nabla_{k_{n+1}}  \cdots \nabla_{k_m}f.
\end{equation}
for $m \ge 1$.
\\

Carce3. If $\sum_{k=a+1}^N \beta^n_k=0$ the R.H.S of (\ref{dddg}) is written as
\begin{equation}\label{S2}
\sum _{m=1}^{\beta^n_a}\left(D^{\bar{a}} \right)^{m-1}\left[D^{\bar{a}},\nabla_{\bar{i}}  \right] \left(D^{\bar{a}} \right)^{\beta^n_a-m}g
=\sum _{m=1}^{\beta^n_a}\sum_{n_a=1}^{\beta^n_a-m} \left(D^{\bar{a}} \right)^{m-1} {{R_{\bar{i}}}^{\bar{a}\bar{a}}}_{\bar{c}}\;  D^{\bar{c}}
\left(D^{\bar{a}} \right)^{\beta^n_a-m-1} g .
\end{equation}

Putting Carce 1,2 and 3 into a shape and recalling that $M$ is a locally symmetric K\"ahler manifold , (\ref{dddg}) is rewritten as
\begin{align}
&\frac{\beta^n_a\left(\beta^n_a-1 \right)}{2}{{R_{\bar{i}}}^{\bar{a}\bar{a}}}_{\bar{c}}\;  D^{\bar{c}}
\left(D^{\bar{a}} \right)^{\beta^n_a-2} \left(D^{\overline{a+1}} \right)^{\beta^n_{a+1}}\cdots \left(D^{\bar{N}} \right)^{\beta^n_N}g   \nonumber  \\
& \qquad +  \sum_{j=a+1}^N \beta^n_j \beta^n_a{{R_{\bar{i}}}^{\bar{a}\bar{j}}}_{\bar{c}}\; D^{\bar{c}}
\left(D^{\bar{a}} \right)^{\beta^n_a-1} \left(D^{\overline{a+1}} \right)^{\beta^n_{a+1}}
\cdots \left(D^{\bar{j}} \right)^{\beta^n_j-1}\cdots \left(D^{\bar{N}} \right)^{\beta^n_N} g. \label{tuika2}
\end{align}


Then we find that the first term of (\ref{lfd1}) is expressed as
\begin{align}
&\hbar\sum_{n=0}^\infty \sum_{\vec{\alpha_n} \vec{\beta_n^*}} \sum _{k=1}^N
  \frac{\beta^n_k\left(\beta^n_k-1 \right)}{2}{{R_{\bar{c}}}^{\bar{k}\bar{k}}}_{\bar{i}}\; T^n_{\vec{\alpha_n}\vec{\beta_n^*}}
\left(D^{\vec{\alpha_n}}f \right)\left(  D^{\vec{\beta_n^*}+\vec{e_c}-2\vec{e_k}}g \right) \notag \\
  &\qquad +\hbar\sum_{n=0}^\infty \sum_{\vec{\alpha_n} \vec{\beta_n^*}} \sum _{k=1}^N\sum _{l=1}^{N-k}\beta^n_k\beta^n_{k+l} {{R_{\bar{c}}}^{\overline{k+l}\bar{k}}}_{\bar{i}}\;
  T^n_{\vec{\alpha_n}\vec{\beta_n^*}} \left( D^{\vec{\alpha_n}}f \right) \left( D^{\vec{\beta_n^*}+\vec{e_c}-\vec{e_k}-\vec{e}_{k+l}}g \right) \label{1stTerm}
\end{align}

Finally , we get the result with substituting (\ref{ddf}) and (\ref{1stTerm}) into (\ref{lfd1})
\begin{align*}
[L_f,\hbar \partial_{\bar{i}}]g
  &=\hbar\sum_{n=0}^\infty \sum_{\vec{\alpha_n} \vec{\beta_n^*}} \sum _{k=1}^N
  \frac{\beta^n_k\left(\beta^n_k-1 \right)}{2}{{R_{\bar{c}}}^{\bar{k}\bar{k}}}_{\bar{i}}\; T^n_{\vec{\alpha_n}\vec{\beta_n^*}}
\left(D^{\vec{\alpha_n}}f \right)\left(  D^{\vec{\beta_n^*}+\vec{e_c}-2\vec{e_k}}g \right)\\
  &\qquad +\hbar\sum_{n=0}^\infty \sum_{\vec{\alpha_n} \vec{\beta_n^*}} \sum _{k=1}^N\sum _{l=1}^{N-k}\beta^n_k\beta^n_{k+l} {{R_{\bar{c}}}^{\overline{k+l}\bar{k}}}_{\bar{i}}\;
  T^n_{\vec{\alpha_n}\vec{\beta_n^*}} \left( D^{\vec{\alpha_n}}f \right) \left( D^{\vec{\beta_n^*}+\vec{e_c}-\vec{e_k}-\vec{e}_{k+l}}g \right)\\
&\qquad -\hbar\sum_{n=0}^\infty \sum_{\vec{\alpha_n} \vec{\beta_n^*}} 
\sum _{d=1}^N g_{\bar{i}d}T^n_{\vec{\alpha_n}\vec{\beta_n^*}}\left( D^{\vec{\alpha_n}+\vec{e_d}}f \right)\left( D^{\vec{\beta_n^*}}g  \right) .
\end{align*}
\qed\end{pf}

\begin{thm}\label{main}
When the star product with separation of variables for smooth functions $f$ and $g$ on a locally symmetric K\"ahler manifold is given as
\[
f * g =\sum_{n=0}^\infty \sum_{\substack{\vec{\alpha_n} \vec{\beta_n^*}  \\ }}
T^n_{\vec{\alpha_n}\vec{\beta_n^*}} \left(D^{\vec{\alpha_n}} f\right) \left(D^{\vec{\beta_n^*}}g\right),
\]
these covariantly constants $T^n_{\vec{\alpha_n}\vec{\beta_n^*}}$ are determined by the following recurrence relations
\begin{align*}
\lefteqn{\sum_{n=0}^\infty\sum_{\substack{\vec{\alpha_n} \vec{\beta_n^*}  \\ }}
 \beta^n_iT^n_{\vec{\alpha_n}\vec{\beta_n^*}} \left(D^{\vec{\alpha_n}} f\right)  \left(D^{\vec{\beta_n^*}-\vec{e_i}}g \right) 
  -\hbar\sum_{n=0}^\infty\sum_{\substack{\vec{\alpha_n} \vec{\beta_n^*}  \\ }} g_{\bar{i}d}
  T^n_{\vec{\alpha_n}\vec{\beta_n^*}}\left( D^{\vec{\alpha_n}+\vec{e_d}}f \right)\left( D^{\vec{\beta_n^*}}g  \right)} \\
&\qquad  +\hbar\sum_{n=0}^\infty\sum_{\substack{\vec{\alpha_n} \vec{\beta_n^*}  \\ }} \sum _{k=1}^N \sum _{p=1}^N\frac{\beta^n_k\left(\beta^n_k-1 \right)}{2}
{{R_{\bar{p}}}^{\bar{k}\bar{k}}}_{\bar{i}}\; T^n_{\vec{\alpha_n}\vec{\beta_n^*}}
\left(D^{\vec{\alpha_n}}f \right)\left(  D^{\vec{\beta_n^*}+\vec{e_p}-2\vec{e_k}}g \right)\\
  &\qquad +\hbar\sum_{n=0}^\infty \sum_{\substack{\vec{\alpha_n} \vec{\beta_n^*}  \\ }} \sum _{\rho=1}^N\sum _{k=1}^{N-1}\sum _{l=1}^{N-k}\beta^n_k\beta^n_{k+l}
   {{R_{\bar{\rho}}}^{\overline{k+l}\bar{k}}}_{\bar{i}}\;
  T^n_{\vec{\alpha_n}\vec{\beta_n^*}} \left( D^{\vec{\alpha_n}}f \right) \left( D^{\vec{\beta_n^*}+\vec{e_\rho}-\vec{e_k}-\vec{e_{k+l}}}g \right) \\
 &=0
\end{align*}
\end{thm}
\begin{pf} $0=\left[L_f, \partial_{\bar{i}}\Phi+\hbar \partial_{\bar{i}} \right]g$ 
is the condition that determines the star product.
$\left[L_f, \partial_{\bar{i}}\Phi\right]g$ and $\left[L_f,\hbar \partial_{\bar{i}} \right]g$ were calculated in Proposition\ref{phi} and \ref{partial}.
\qed\end{pf}

\begin{thm}\label{3_6}
When the star product with separation of variables for smooth functions $f$ and $g$ on a local symmetric K\"ahler manifold is given as
\[
f * g =\sum_{n=0}^\infty \sum_{\substack{\vec{\alpha_n} \vec{\beta_n^*}  \\ }}
T^n_{\vec{\alpha_n}\vec{\beta_n^*}} \left(D^{\vec{\alpha_n}} f\right) \left(D^{\vec{\beta_n^*}}g\right),
\]
these smooth functions $T^n_{\vec{\alpha_n}\vec{\beta_n^*}}$, which are covariantly constants, 
are determined by the following recurrence relations for $\forall i:$
\begin{align*}
\lefteqn{\sum _{d=1}^N\hbar g_{\bar{i}d}T^{n-1}_{\vec{\alpha_n}-\vec{e_d}\vec{\beta_n^*}-\vec{e_i}} } \\
 &=\beta^n_iT^n_{\vec{\alpha_n}\vec{\beta_n^*}}
 +\sum _{k=1}^N\sum _{p=1}^N\frac{\hbar\left(\beta^n_k-\delta_{kp}-\delta_{ik} +1 \right)\left(\beta^n_k-\delta_{kp}-\delta_{ik}+2 \right)}{2}
{{R_{\bar{p}}}^{\bar{k}\bar{k}}}_{\bar{i}}\; T^n_{\vec{\alpha_n}\vec{\beta_n^*}-\vec{e_p}+2\vec{e_k}-\vec{e_i}}\\
  &\qquad + \sum _{k=1}^{N-1}\sum _{l=1}^{N-k}\sum _{p=1}^N
  \hbar\left(\beta^n_k-\delta_{kp}-\delta_{ik}+1 \right)\left(\beta^n_{k+l}-\delta_{\left( k+l\right),p}-\delta_{i,\left( k+l\right)}+1  \right)
  {{R_{\bar{p}}}^{\overline{k+l}\bar{k}}}_{\bar{i}}\; T^n_{\vec{\alpha_n}\vec{\beta_n^*}-\vec{e_p}+\vec{e_k}+\vec{e_{k+l}}-\vec{e_i}}.
\end{align*}
\end{thm}

\begin{pf} Changing the summation of Theorem \ref{main},
\begin{align*}
\lefteqn{\hbar\sum_{n=1}^\infty\sum_{\vec{\alpha_n} \vec{\beta_n^*}} \sum _{d=1}^N g_{\bar{i}d}
T^{n-1}_{\vec{\alpha_n-\vec{e_d}}\vec{\beta_n^*}-\vec{e_i}}\left( D^{\vec{\alpha_n}}f \right)\left( D^{\vec{\beta_n^*}-\vec{e_i}}g  \right) } \\
 &=\sum_{n=0}^\infty\sum_{\vec{\alpha_n} \vec{\beta_n^*}} \beta^n_iT^n_{\vec{\alpha_n}\vec{\beta_n^*}} \left(D^{\vec{\alpha_n}} f\right)  \left(D^{\vec{\beta_n^*}-\vec{e_i}}g \right) \\
&\qquad  +\hbar\sum_{n=0}^\infty \sum _{k=1}^N\sum _{p=1}^N\sum_{\vec{\alpha_n} \vec{\beta_n^*}}
\frac{\left(\beta^n_k-\delta_{kp}-\delta_{ik} +1 \right)\left(\beta^n_k-\delta_{kp}-\delta_{ik}+2 \right)}{2}
{{R_{\bar{p}}}^{\bar{k}\bar{k}}}_{\bar{i}}\;\\
 &\qquad  \qquad  \times T^n_{\vec{\alpha_n}\vec{\beta_n^*}-\vec{e_p}+2\vec{e_k}-\vec{e_i}} 
  \left(D^{\vec{\alpha_n}}f \right)\left(  D^{\vec{\beta_n^*}-\vec{e_i}}g \right)\\
  &\qquad +\hbar\sum_{n=0}^\infty \sum _{k=1}^{N-1}\sum _{l=1}^{N-k}\sum _{p=1}^N\sum_{\vec{\alpha_n} \vec{\beta_n^*}}
  \left(\beta^n_k-\delta_{kp}-\delta_{ik}+1 \right)\left(\beta^n_{k+l}-\delta_{\left( k+l\right),p}-\delta_{i,\left( k+l\right)}+1  \right)
  {{R_{\bar{p}}}^{\overline{k+l}\bar{k}}}_{\bar{i}}\; \\
 &\qquad  \qquad \times T^n_{\vec{\alpha_n}\vec{\beta_n^*}-\vec{e_p}+\vec{e_k}+\vec{e_{k+l}}-\vec{e_i}} \left( D^{\vec{\alpha_n}}f \right) \left( D^{\vec{\beta_n^*}-\vec{e_i}}g \right) ,
\end{align*}
and this implies the theorem.
\qed\end{pf}


\section{One and two dimensional cases}
\label{one}

By using Theorem \ref{3_6}
we can provide explicit star products for locally symmetric K\"ahler manifolds.
In this section, an explicit expression of a star product of a one-dimensional locally symmetric K\"ahler manifold is constructed as an example.
A two-dimensional locally symmetric K\"ahler manifold is also considered.\\

At first, we study an explicit expression of 
a star product of a one-dimensional
locally symmetric K\"ahler manifold.
A formal discussions are given in \cite{Schlichenmaier2}, and 
star products are studied in \cite{Ohsaku:2006am}.
Complex surfaces with arbitrary genus 
are known as a example of such manifolds when we chose proper coordinates
and metrics.
The Scalar curvature $R$ is defined as
\[
R=g^{i\bar{j}}R_{i\bar{j}}
={{R_{\bar{l}}}^{\bar{j}\bar{l}}}_{\bar{j}}.
\]

\begin{prop}Let $M$ be a one-dimensional locally symmetric K\"ahler manifold $\left(N=1 \right)$ and $f$ and $g$ be smooth functions on $M$.
The star product with separation of variables for $f$ and $g$ can be described as
\[
f * g =\sum_{n=0}^\infty 
\left[\left(g_{1\bar{1}} \right)^n\left\{\prod_{k=1}^{n-1} \frac{2\hbar }{2k+\hbar k\left(k-1 \right)R}\; \right\}
\left\{\left(g^{1\bar{1}} \frac{\partial}{\partial z} \right)^nf \right\}
\left\{\left(g^{1\bar{1}}\frac{\partial}{\partial\bar{z}}  \right)^ng \right\} \right]
\]
where
\[
R={{R_{\bar{1}}}^{\bar{1}\bar{1}}}_{\bar{1}}.
\]
\end{prop}
\begin{pf} $N=1, i=1$ and  
\[
D^{\vec{\alpha_n}} f=\left(g^{1\bar{1}} \frac{\partial}{\partial z} \right)^nf,\qquad  D^{\vec{\beta_n^*}}g=\left(g^{1\bar{1}}\frac{\partial}{\partial\bar{z}}  \right)^ng
\]
are substituted in Theorem \ref{main} , then we obtain
\[
\hbar\sum_{n=1}^\infty g_{1\bar{1}}T^{n-1}\left( D^nf_1 \right)\left( D^{n-1}f_2  \right) 
=\sum_{n=0}^\infty \left\{n+\frac{\hbar n\left(n-1 \right)}{2}{{R_{\bar{1}}}^{\bar{1}\bar{1}}}_{\bar{1}} \right\}
 T^n\left(D^nf_1 \right)\left(  D^{n-1}f_2 \right)
\]
or equivalently,
the recurrence relation of $T^n$ is given as
\[
T^n=  g_{1\bar{1}}\left\{\frac{2\hbar } {2n+\hbar n\left(n-1 \right)R}\right\}T^{n-1}.
\]
From Proposition \ref{t1} the first term $T^1$ is given as
$
T^1=\hbar g^{1\bar{1}}$.
Then, $T^n$ is given as
\[
T^n=\left(g_{1\bar{1}} \right)^n\prod_{k=1}^{n-1} \left\{\frac{2\hbar }{2k+\hbar k\left(k-1 \right)R} \right\}.
\]
\qed\end{pf}

Next, we discuss star products on general two-dimensional locally symmetric K\"ahler manifolds.

According to Proposition \ref{t1}, for a two-dimensional locally symmetric K\"ahler manifold 
$M$, $T^1_{\vec{\alpha_1}\vec{\beta_1^*}}$ is given as
\[
\left(
\begin{array}{cc}
T^1_{\left(1,0\right),\left(1,0 \right)} & T^1_{\left(1,0\right),\left(0,1 \right)}\\
T^1_{\left(0,1\right),\left(1,0 \right)} & T^1_{\left(0,1\right),\left(0,1 \right)}
\end{array}
\right)=
\hbar \left(
\begin{array}{cc}
g_{1\bar{1}} & g_{1\bar{2}}\\
g_{2\bar{1}} & g_{2\bar{2}}
\end{array}
\right).
\]
Next, we estimate $T^2_{\vec{\alpha_2}\vec{\beta_2^*}}$.
\begin{prop}Let $M$ be a two-dimensional locally symmetric K\"ahler manifold  and $f$ and $g$ be smooth functions on $M$.
$T^2_{\vec{\alpha_2}\vec{\beta_2^*}}$ given in (\ref{lfg}) is obtained by \label{2dim}
\begin{align*}
\lefteqn{\left(
\begin{array}{ccc}
T^2_{\left(2,0\right),\left(2,0 \right)} & T^2_{\left(2,0\right),\left(1,1 \right)}& T^2_{\left(2,0\right),\left(0,2 \right)}\\
T^2_{\left(1,1\right),\left(2,0 \right)} & T^2_{\left(1,1\right),\left(1,1 \right)}& T^2_{\left(1,1\right),\left(0,2 \right)}\\
T^2_{\left(0,2\right),\left(2,0 \right)} & T^2_{\left(0,2\right),\left(1,1 \right)}& T^2_{\left(0,2\right),\left(0,2 \right)}
\end{array}
\right)} \\
 &= \hbar^2\left(
\begin{array}{ccc}
\left(g_{\bar{1}1} \right)^2  & g_{\bar{1}1}g_{\bar{2}1}  &  \left(g_{\bar{2}1} \right)^2    \\
 2g_{\bar{1}1} g_{\bar{1}2} & g_{\bar{2}1}g_{\bar{1}2}+ g_{\bar{1}1}g_{\bar{2}2}& 2 g_{\bar{2}1} g_{\bar{2}2} \\
\left(g_{\bar{1}2} \right)^2  & g_{\bar{2}1}g_{\bar{2}2}  &  \left(g_{\bar{2}2} \right)^2
\end{array}
\right)
\left(
\begin{array}{ccc}
2+\hbar{{R_{\bar{1}}}^{\bar{1}\bar{1}}}_{\bar{1}}  & \hbar {{R_{\bar{2}}}^{\bar{1}\bar{1}}}_{\bar{1}}  
& \hbar {{R_{\bar{2}}}^{\bar{1}\bar{1}}}_{\bar{2}} \\
\hbar {{R_{\bar{1}}}^{\overline{2}\bar{1}}}_{\bar{1}} &1+\hbar {{R_{\bar{2}}}^{\overline{2}\bar{1}}}_{\bar{1}}  
& \hbar {{R_{\bar{2}}}^{\overline{2}\bar{1}}}_{\bar{2}}\\
 \hbar {{R_{\bar{1}}}^{\bar{2}\bar{2}}}_{\bar{1}} &\hbar {{R_{\bar{2}}}^{\bar{2}\bar{2}}}_{\bar{1}} 
 & 2+\hbar {{R_{\bar{2}}}^{\bar{2}\bar{2}}}_{\bar{2}} 
\end{array}
\right)^{-1}.
\end{align*}
\end{prop}

The proof is given in appendix \ref{aa}.


\section{Deformation quantization for complex Grassmann manifold}
\label{grassmann}

In this section, recurrence relations to obtain star products on complex Grassmann manifolds are derived.
Especially we calculate star products of ${\mathbb C}P^N$.
Note that this star product is also equal to the ones given in 
\cite{Bordemann,Hayasaka:2002db,Sako:2013noa}, 
and if we put a some restriction 
our star product is also equal to the one given in \cite{Balachandran},
as they are shown in \cite{Sako:2012ws,Maeda:2015bnb}.
The equivalence is also discussed in \cite{Schlichenmaier1,Schlichenmaier2}.
In addition, recurrence relations to construct star products for $G_{2,2}$ 
was derived. 
Deformation quantization of Grassmann manifolds and flag manifolds were 
studied in 
\cite{Karabegov:1997wd,Majdi Ben Halima,Majdi Ben Halima2,Murray:2006pi}.

Complex Grassmann manifold $G_{p,q}$ is defined as a set of the whole $p$ dimensional part vector space of $p+q$ dimensional vector space $V$.
The local coordinate can be defined in a similar way to S. Kobayashi and K. Nomizu pp.160-162\cite{Kobayashi_Nomizu}.

Let $U$ be an open subset of $G_{p,q}$ . A chart $\left(U,\phi \right)$ is defined by 
\[
U:=\left\{ Y=\left(
\begin{array}{c}
 Y_0 \\
 Y_1
\end{array}
\right)\in M\left(p+q,p;{\mathbb C} \right)
;\left|Y_0 \right|\neq 0\right\}
\]
and
\[
\phi:U\longrightarrow M\left(q,p;{\mathbb C} \right)
\]
where
\[
\phi\left(Y \right)=Y_1Y_0^{-1}.
\]
This is a holomorphic map of $U$ onto an open subset of $p\times q$-dimensional complex space. 

In this section, capital letter indices $A,B,C\cdots $ mean $aa',bb',cc'\cdots $.
In the inhomogeneous coordinates $z^I:=z^{ii'},~ z^{\bar{I}}:=z^{\bar{i}\bar{i'}},(i=1,2,\cdots p,i'=1,2,\cdots q)$
, the K\"ahler potential of $G_{p,q}$ is given as
\begin{align}
 \Phi = \ln \left|E_q+Z^\dagger Z  \right|, \label{kp}
\end{align}
where $Z=\phi\left(Y \right)=(z^{I}) \in M\left(q,p;{\mathbb C} \right)$
and $E_q \in M\left(q,q;{\mathbb C} \right)$ is the unite matrix.
{} From (\ref{kp}), the following facts are derived.
\begin{fact}\label{metric}
The Fubini-Study metric $(g_{I\bar{J}})$ is 
\[
 ds^2 = 2g_{I\bar{J}}dz^{I}d\bar{z}^{J}, \label{ds} 
 \]
 where
\[
g_{I\bar{J}} :=g_{ii'\bar{j}\bar{j'}}= \partial _{I}\partial _{\bar{J}} \Phi  = a^{ji}b^{i'j'}
,\: g^{I\bar{J}} := g^{ii'\bar{j}\bar{j'}} = a_{ij}b_{j'i'}.
\]
with
\[
a_{ij}=\delta_{ij}+z^{ik'}\bar{z}^{jk'},\: b_{i'j'}=\delta_{i'j'}+\bar{z}^{ki'} z^{kj'}.
\]
\end{fact}
\begin{fact}The curvature of a complex Grassmann manifold is
\begin{align}
\label{tuika3}
{{R_{\bar{A}}}^{\bar{C}\bar{D}}}_{\bar{B}}=g^{P\bar{C}}g^{Q\bar{D}}R_{\bar{A}PQ\bar{B}}
=-\delta_{\overline{ab'}\:\overline{C}}\delta_{\overline{ba'}\:\overline{D}}-\delta_{\overline{ba'}\:\overline{C}}\delta_{\overline{ab'}\:\overline{D}},
\end{align}
where
\begin{equation*}
\delta_{\overline{ab'}\: \overline{cd'}}= \left \{
\begin{array}{l}
1\qquad (a=c,b'=d') \\
0\qquad (otherwise)
\end{array}
\right. .
\end{equation*}
\end{fact}

From these facts, we can derive the recurrence relations to determine
star products on the Grassmann manifolds.

\subsection{Some preparations}

A function similar to the determinant is defined on the matrix space.
\begin{df}Let $C=\left(C_{k,l} \right)_{1\leq k\leq n,1\leq l\leq n}$ be a $n\times n$ matrix. We define $\left|\: \cdot \:  \right|^+$ 
as a $\mathbb{C}$-valued function on $M\left(n,n;{\mathbb C} \right)$ 
such that
\[
\left|C \right|^+:=\sum_{\sigma_n\in S_n}\prod_{k=1}^n C_{k,\sigma_n\left(k \right)}.
\]
\end{df}

\begin{ex}Here we show some examples. These suggest some properties like determinant.
\begin{enumerate}
 \item \[
\left|
\begin{array}{cc}
c_{11} &c_{12}  \\
c_{21} &c_{22}
\end{array}
\right|^+ =c_{11}c_{22}+c_{12}c_{21}
 \]
  \item 
\begin{align*}
\lefteqn{\left|
\begin{array}{ccc}
c_{11} &c_{12}  & c_{13} \\
c_{21} &c_{22}  & c_{23}  \\
c_{31} &c_{32}  & c_{33} 
\end{array}
\right|^+} \\
 &=c_{11}c_{22}c_{33}+c_{11}c_{23}c_{32}+c_{12}c_{21}c_{33}+c_{12}c_{23}c_{31}+c_{13}c_{21}c_{32}+c_{13}c_{22}c_{31} \\
 &=c_{11}\left|
\begin{array}{cc}
c_{22}  & c_{23}  \\
c_{32}  & c_{33} 
\end{array}
\right|^+
+c_{12}\left|
\begin{array}{cc}
c_{11} & c_{13} \\
c_{31} & c_{33} 
\end{array}
\right|^+
+c_{13}\left|
\begin{array}{ccc}
c_{11} &c_{12} \\
c_{21} &c_{22}
\end{array}
\right|^+
\end{align*}
\end{enumerate}
\end{ex}

\begin{rem}Similar to a determinant
\[
\left|^tC \right|^+=\left|C \right|^+,
\]
where $^tC$ is a transposed matrix of $C$.
\end{rem}

The following is a proposition similar to cofactor expansion of a determinant.
\begin{prop}\label{cof}
\[
\left|C \right|^+=\left|
\begin{array}{ccccc}
c_{11} &\cdots &c_{1j} &\cdots & c_{1n} \\
 \vdots   &\ddots  &\vdots  &\ddots  & \vdots  \\
c_{i1} &\cdots &c_{ij} &\cdots & c_{in}  \\
 \vdots   &\ddots  &\vdots  &\ddots  & \vdots   \\
c_{n1} &\cdots  &c_{nj} &\cdots   & c_{nn} 
\end{array}
\right|^+
=\sum_{j=1}^nc_{ij}\left|
\begin{array}{ccccc}
c_{11} &\cdots &\hat{c_{1j}} &\cdots & c_{1n} \\
 \vdots   &\ddots  &\vdots  &\ddots  & \vdots  \\
\hat{c_{i1}} &\cdots &\hat{c_{ij}} &\cdots & \hat{c_{in}}  \\
 \vdots   &\ddots  &\vdots  &\ddots  & \vdots   \\
c_{n1} &\cdots  &\hat{c_{nj}} &\cdots   & c_{nn} 
\end{array}
\right|^+
\]
\end{prop}
\begin{pf} A proof for this function is similar to the case of determinants.
\qed\end{pf}

\begin{df} \label{tuika6}
A matrix $G^{\vec{\alpha_n},\vec{\beta^*_n}}$ is defined by using the Riemannian metrics on $M$.
Its elements are metrics on $M$ and are located  as follows.
$\vec{\alpha_n}$ and $\vec{\beta_n}$ are elements of $\mathbb{Z}^N$.

\[
G^{\vec{\alpha_n},\vec{\beta^*_n}}=
\left(
\begin{array}{ccc}
\tilde{G}_{11} &\cdots  & \tilde{G}_{1n} \\
\vdots  &\ddots  & \vdots  \\
\tilde{G}_{n1} &\cdots  & \tilde{G}_{nn}
\end{array}
\right)
\]
where
\[
\tilde{G}_{pq}=:g_{p\bar{q}} 
\left(
\begin{array}{ccc}
1&\cdots  &1 \\
\vdots  &\ddots  & \vdots  \\
1&\cdots  &1
\end{array}
\right)
\in M\left(\alpha_p^n,\beta_q^n;{\mathbb C} \right)
\]
i.e.

\[
G^{\vec{\alpha_n},\vec{\beta^*_n}}=
\begin{array}{rccc|c|ccccll}
\ldelim({9}{4pt}[] &g_{1\bar{1}}     &\cdots&g_{1\bar{1}} &  &g_{1\bar{N}}  &\cdots &g_{1\bar{N}}     &\rdelim){9}{4pt}[]&\rdelim\}{4}{10pt}[$\alpha_1^n$] \\
                   &\vdots     &\ddots &\vdots &\cdots   &\vdots&\ddots& \vdots  &\\
                   &g_{1\bar{1}}&\cdots &g_{1\bar{1}}& & g_{1\bar{N}}    &\cdots &g_{1\bar{N}}  & &\\ \cline{2-8}
                   &   &\vdots  &  &\ddots  &   & \vdots   &  &&\vdots \\ \cline{2-8}
                   &g_{N\bar{1}}     &\cdots&g_{N\bar{1}} &      &g_{N\bar{N}}  &\cdots &g_{N\bar{N}}     &&\rdelim\}{4}{10pt}[$\alpha_N^n$] \\
                   &\vdots     &\ddots &\vdots &\cdots  &\vdots&\ddots& \vdots  &\\
                   &g_{N\bar{1}}&\cdots &g_{N\bar{1}}& & g_{N\bar{N}}    &\cdots &g_{N\bar{N}}  &                  &\\
                   &\multicolumn{8}{l}{\underbrace{\hspace{7em}}_{\mbox{$\beta_1^n$}}\ \ \cdots \ \ \underbrace{\hspace{6em}}_{\mbox{$\beta_N^n$}}} &
\end{array} .
\]

\end{df}

For example $N=2,\vec{\alpha_3}=\left(2,1 \right),\vec{\beta_3^*}=\left(1,2 \right)^*$, then $G^{\vec{\alpha_3},\vec{\beta_3^*}}$ is determined as
\begin{align*}
G^{\vec{\alpha_3},\vec{\beta_3^*}}=\left(
\begin{array}{c|cc}
g_{1\bar{1}} &g_{1\bar{2}} & g_{1\bar{2}} \\
g_{1\bar{1}} &g_{1\bar{2}} & g_{1\bar{2}}  \\
\hline
g_{2\bar{1}} &g_{2\bar{2}} & g_{2\bar{2}} 
\end{array}
\right).
\end{align*}


From Proposition \ref{cof}, we obtain the following corollary.
\begin{cor}\label{gab}For a matrix $G^{\vec{\alpha_n},\vec{\beta^*_n}} $,
\[
\left|G^{\vec{\alpha_n},\vec{\beta^*_n}} \right|^+=\sum_{J=1}^N\beta^n_Jg_{\bar{J}I}\left|G^{\vec{\alpha_n}-\vec{e_I},\vec{\beta^*_n}-\vec{e_J}} \right|^+
 =\sum_{K=1}^N\alpha^n_Kg_{\bar{I}K}\left|G^{\vec{\alpha_n}-\vec{e_K},\vec{\beta^*_n}-\vec{e_I}} \right|^+.
\]
\end{cor}


\subsection{Deformation quantization for a complex projective space}
In this subsection, we obtain concrete expression of star products 
on ${\mathbb C}P^N$.
A complex projective space ${\mathbb C}P^N$ is a Grassmann manifold $G_{1,N}$ by definition.

\begin{prop}
Let $M$ be a complex projective space and $f$ and $g$ be smooth functions on $M$.\label{projective}
The recurrence relation of $\: T^n_{\vec{\alpha_n}\vec{\beta_n^*}}$ given in (\ref{lfg}) is
\begin{equation} \label{cp}
T^n_{\vec{\alpha_n}\vec{\beta^*_n}} 
= \sum_{d=1}^N \frac{\hbar g_{\bar{i}d}}{\left( 1+\hbar -\hbar n \right)\beta^n_i}T^{n-1}_{\vec{\alpha_n}-\vec{e_d}\vec{\beta^*_n}-\vec{e_i}}  .
\end{equation}
\end{prop}
\begin{pf}
The curvature (\ref{tuika3}) is substituted for Theorem \ref{3_6}, and the following is proved.
\begin{align*}
\lefteqn{\sum _{d=1}^N\hbar g_{\bar{i}d}T^{n-1}_{\vec{\alpha_n}-\vec{e_d}\vec{\beta_n^*}-\vec{e_i}} } \\
 &=\beta_i^nT^n_{\vec{\alpha_n}\vec{\beta_n^*}}
 +\sum _{k=1}^N\sum _{p=1}^N\frac{\hbar\left(\beta^n_k-\delta_{kp}-\delta_{ik} +1 \right)\left(\beta^n_k-\delta_{kp}-\delta_{ik}+2 \right)}{2}
{{R_{\bar{p}}}^{\bar{k}\bar{k}}}_{\bar{i}}\; T^n_{\vec{\alpha_n}\vec{\beta_n^*}-\vec{e_p}+2\vec{e_k}-\vec{e_i}}\\
  &\qquad + \sum _{k=1}^{N-1}\sum _{l=1}^{N-k}\sum _{p=1}^N
  \hbar\left(\beta^n_k-\delta_{kp}-\delta_{ik}+1 \right)\left(\beta^n_{k+l}-\delta_{\left( k+l\right),p}-\delta_{i,\left( k+l\right)}+1  \right)
  {{R_{\bar{p}}}^{\overline{k+l}\bar{k}}}_{\bar{i}}\; T^n_{\vec{\alpha_n}\vec{\beta_n^*}-\vec{e_p}+\vec{e_k}+\vec{e_{k+l}}-\vec{e_i}}.
\end{align*}
We also use
\[
{{R_{\bar{p}}}^{\overline{k+l}\bar{k}}}_{\bar{i}}
=-\delta_{\overline{p},\overline{k+l}}\delta_{\overline{i},\bar{k}}-\delta_{\overline{i},\overline{k+l}}\delta_{\overline{p},\bar{k}}
,{{R_{\bar{p}}}^{\bar{k}\bar{k}}}_{\bar{i}}
=-\delta_{\overline{p},\bar{k}}\delta_{\overline{i},\bar{k}}-\delta_{\overline{i},\bar{k}}\delta_{\overline{p},\bar{k}} ,
\]
then the above is rewritten as
\begin{align*}
\lefteqn{\sum _{d=1}^N\hbar g_{\bar{i}d}T^{n-1}_{\vec{\alpha_n}-\vec{e_d}\vec{\beta_n^*}-\vec{e_i}} } \\
 &=\beta_i^nT^n_{\vec{\alpha_n}\vec{\beta_n^*}}
 -\sum _{k=1}^N\sum _{p=1}^N\frac{\hbar\left(\beta^n_k-\delta_{kp}-\delta_{ik} +1 \right)\left(\beta^n_k-\delta_{kp}-\delta_{ik}+2 \right)
 \left(\delta_{\overline{p},\bar{k}}\delta_{\overline{i},\bar{k}}+\delta_{\overline{i},\bar{k}}\delta_{\overline{p},\bar{k}} \right)}{2}
\; T^n_{\vec{\alpha_n}\vec{\beta_n^*}-\vec{e_p}+2\vec{e_k}-\vec{e_i}} \\
 &\qquad - \sum _{k=1}^{N-1}\sum _{l=1}^{N-k}\sum _{p=1}^N
  \hbar\left(\beta^n_k-\delta_{kp}-\delta_{ik}+1 \right)\left(\beta^n_{k+l}-\delta_{\left( k+l\right),p}-\delta_{i,\left( k+l\right)}+1  \right) \\
 &\qquad \qquad \times   \left(\delta_{\overline{p},\overline{k+l}}\delta_{\overline{i},\bar{k}}+\delta_{\overline{i},\overline{k+l}}\delta_{\overline{p},\bar{k}} \right)
  \; T^n_{\vec{\alpha_n}\vec{\beta_n^*}-\vec{e_p}+\vec{e_k}+\vec{e_{k+l}}-\vec{e_i}}.
\end{align*}

The theorem follows from this.
\qed\end{pf}

\begin{thm}\label{cpthm}Let $f$ and $g$ be smooth functions on a projective space ${\mathbb C}P^N$.
A star product with separation of variables on a projective space ${\mathbb C}P^N$ is given as
\begin{align}
f * g =f\cdot g+\sum_{n=1}^\infty \sum_{\substack{\vec{\alpha_n} \vec{\beta^*_n}  \\ }}
 \left|G^{\vec{\alpha_n},\vec{\beta^*_n}} \right|^+\left\{\prod_{k=0}^n\frac{\hbar}
{\left(1+\hbar-\hbar k \right)\alpha^n_k!\beta^n_k!}\right\} \left(D^{\vec{\alpha_n}} f\right) \left(D^{\vec{\beta^*_n}}g\right). \label{tuika4}
\end{align}
\end{thm}
\begin{pf}
We show that 
\[T^n_{\vec{\alpha_n}\vec{\beta^*_n}}=
 \left|G^{\vec{\alpha_n},\vec{\beta^*_n}} \right|^+\left\{\prod_{k=0}^n\frac{\hbar}
{\left(1+\hbar-\hbar k \right)\alpha^n_k!\beta^n_k!}\right\}
\]
satisfies (\ref{cp}).
The R.H.S of (\ref{cp}) for this case is given as
\[
\sum_{d=1}^N\frac{\hbar g_{\bar{i}d} }{\left( 1+\hbar-\hbar n \right)\beta^n_i}T^{n-1}_{\vec{\alpha_n}-\vec{e_d}\vec{\beta^*_n}-\vec{e_i}} 
= \sum_{d=1}^Ng_{\bar{i}d}\alpha^n_d\left|G^{\vec{\alpha_n}-\vec{e_d},\vec{\beta^*_n}-\vec{e_i}} 
\right|^+ \frac{\hbar} {\left( 1+\hbar-\hbar n \right)}
 \prod_{k=0}^{n-1}\frac{\hbar}{\left(1+\hbar-\hbar k \right)\alpha^n_k!\beta^n_k!}.
\]
Using Corollary \ref{gab}, 
R.H.S. of the above is rewritten as
\[
\left|G^{\vec{\alpha_n},\vec{\beta^*_n}} \right|^+\prod_{k=0}^n\frac{\hbar}{\left(1+\hbar-\hbar k \right)\alpha^n_k!\beta^n_k!}.
\]
This shows the given $T^n_{\vec{\alpha_n}\vec{\beta^*_n}}$ satisfies the recurrence relation (\ref{cp}).
\qed\end{pf}


\begin{fact}\label{2013noa} Let $f$ and $g$ be smooth functions on a projective space ${\mathbb C}P^N$.
A star product on a projective space ${\mathbb C}P^N$ is given in \cite{Sako:2013noa} as
\begin{align}
\lefteqn{f \tilde{*} g = \sum_{n=0}^\infty \frac{\Gamma(1-n+1/\hbar) g^{\bar{j}_1 k_1} \cdots g^{\bar{j}_n k_n} }{n! \Gamma(1+1/\hbar)}
 \left(\nabla_{\bar{j}_1} \cdots \nabla_{\bar{j}_n} f\right)
 \left(\nabla_{k_1} \cdots \nabla_{k_n} g\right)} \notag\\
 &=\sum_{n=0}^\infty \frac{\Gamma(1-n+1/\hbar) }{n! \Gamma(1+1/\hbar)}
 \left(D^{k_1} \cdots D^{k_n} f \right) \left(\nabla_{k_1} \cdots \nabla_{k_n} g\right) \notag \\
 &=\sum_{n=0}^\infty \frac{\Gamma(1-n+1/\hbar)g_{\bar{m}_1 k_1} \cdots g_{\bar{m}_n k_n} }{n! \Gamma(1+1/\hbar)}
 \left(D^{k_1} \cdots D^{k_n} f \right) \left( D^{\bar{m}_1} \cdots D^{\bar{m}_n} g\right). \label{tuika5}
\end{align}
\end{fact}
As mentioned in Section \ref{Kahler}, the star product with separation of variables is uniquely determined.
This fact means (\ref{tuika4}) coincides with (\ref{tuika5}).
This coincidence is easily checked from Definition \ref{tuika6}.
\subsection{Deformation quantization for a $G_{2,2}$}
In this subsection, we derive the recurrence relation to 
obtain concrete expression of star products on 
a Grassmann manifold $G_{2,2}$. 
The inhomogeneous coordinates are $z^{11'} , z^{12'} , z^{21'}$ and $ z^{22'}$.
 To decide the order of coordinates is useful in order 
to calculate the finite sum. We set the order:$11'<12'<21'<22'$. 
In this subsection, $j$ is used as ``Not $i$''.
That means that if $i=1$ then $j=2$ and if $i=2$ then $j=1$.
For example, if $I=ii'=11'$, then $ij'=12', ji'=21', J=22'$.
If $I=ii'=12'$, then $ij'=11', ji'=22', J=21'$.
A finite sum is defined as 
\[
\sum _{D=1}^4a_D:=a_{11'}+a_{12'}+a_{21'}+a_{22'}.
\]

\begin{thm}Let  $f$ and $g$ be smooth functions on $G_{2,2}$.\label{G22}
The recurrence relation of $\: T^n_{\vec{\alpha_n}\vec{\beta_n^*}}$ given in (\ref{lfg}) is
\begin{align}
\lefteqn{ \beta_I^n\left(1+\hbar-\hbar\beta^n_I
-\hbar\beta^n_{ji'} -\hbar\beta^n_{ij'}  \right)\; T^n_{\vec{\alpha_n}\vec{\beta_n^*}} 
-\hbar\left(\beta^n_{ij'}+1 \right)  \left(\beta^n_{ji'}+1  \right)
  \; T^n_{\vec{\alpha_n}\vec{\beta_n^*}-\vec{e_J}+\vec{e_{ij'}}+\vec{e_{ji'}}-\vec{e_I}}} \nonumber  \\
 &=\hbar g_{\bar{I}I}T^{n-1}_{\vec{\alpha_n}-\vec{e_I}\vec{\beta_n^*}-\vec{e_I}}
+\hbar g_{\bar{I}ij'}T^{n-1}_{\vec{\alpha_n}-\vec{e_{ij'}}\vec{\beta_n^*}-\vec{e_I}}
+\hbar g_{\bar{I}ji'}T^{n-1}_{\vec{\alpha_n}-\vec{e_{ji'}}\vec{\beta_n^*}-\vec{e_I}}
+\hbar g_{\bar{I}J}T^{n-1}_{\vec{\alpha_n}-\vec{e_J}\vec{\beta_n^*}-\vec{e_I}}.\label{g22}
\end{align}
for each $I$.
\end{thm}
\begin{pf}
The curvature (\ref{tuika3}) is substituted into Theorem \ref{3_6}, and the following is obtained.
\begin{align*}
\lefteqn{\sum _{D=1}^4\hbar g_{\bar{I}D}T^{n-1}_{\vec{\alpha_n}-\vec{e_D}\vec{\beta_n^*}-\vec{e_I}} } \\
 &=\beta_I^nT^n_{\vec{\alpha_n}\vec{\beta_n^*}} \\
 & -\sum _{K=1}^4\sum _{P=1}^4\frac{\hbar\left(\beta^n_K-\delta_{KP}-\delta_{IK} +1 \right)\left(\beta^n_K-\delta_{KP}-\delta_{IK}+2 \right)
 \left(\delta_{\overline{pi'},\bar{K}}\delta_{\overline{ip'},\bar{K}}+\delta_{\overline{ip'},\bar{K}}\delta_{\overline{pi'},\bar{K}} \right)}{2}
\; T^n_{\vec{\alpha_n}\vec{\beta_n^*}-\vec{e_P}+2\vec{e_K}-\vec{e_I}}\\
  &\qquad -\sum _{K=1}^{4-1}\sum _{L=1}^{4-K} \sum _{P=1}^4
  \hbar\left(\beta^n_K-\delta_{KP}-\delta_{IK}+1 \right)\left(\beta^n_{K+L}-\delta_{\left( K+L\right),P}-\delta_{I,\left( K+L\right)}+1  \right) \\
 & \qquad \qquad \times \left(\delta_{\overline{pi'},\overline{K+L}}\delta_{\overline{ip'},\bar{K}}+\delta_{\overline{ip'},\overline{K+L}}\delta_{\overline{pi'},\bar{K}} \right) 
 \; T^n_{\vec{\alpha_n}\vec{\beta_n^*}-\vec{e_P}+\vec{e_K}+\vec{e_{K+L}}-\vec{e_I}} \\
& = \beta_I^n\left\{1+\hbar-\hbar\beta^n_I
-\hbar\beta^n_{ji'} -\hbar\beta^n_{ij'} \right\}\; T^n_{\vec{\alpha_n}\vec{\beta_n^*}} 
-\hbar\left(\beta^n_{ij'}+1 \right)  \left(\beta^n_{ji'}+1  \right)
  \; T^n_{\vec{\alpha_n}\vec{\beta_n^*}-\vec{e_J}+\vec{e_{ij'}}+\vec{e_{ji'}}-\vec{e_I}}
\end{align*}
The theorem follows from this.
\qed\end{pf}

Star products on a noncommutative $G_{2,2}$ are determined 
by this formula recursively.
For general $G_{p,q}$, the recurrnce relations are determined in a similar way.

\section{Conclusion} 
\label{conclusion}

In this article, noncommutative locally symmetric K\"ahler manifolds have been 
studied by using deformation quantization
given by Karabegov \cite{Karabegov1996}, which is called deformation quantization with separation of variables.
The similar approach was already tried in \cite{Sako:2012ws}, and 
recursion relations and explicit expression of 
star products for ${\mathbb C}P^N$ and ${\mathbb C}H^N$ are also given in it.
In this article, we improve the method in \cite{Sako:2012ws} 
to more useful one. The point to obtain the concrete star products is 
to translate the system of PDE into algebraic recursion relations. 
The fact that the Riemannian curvature tensor 
is given as a covariantly constant removes the 
complication of the system of PDE.
{}From the results, we can construct explicit star products order by order 
by using only algebraic calculations.
As an example, a concrete expression of a star product for 
${\rm dim}_{\mathbb C} M =1$ locally symmetric K\"ahler manifolds was
obtained in Section 4.
It is know that Riemann surfaces with arbitrary genus are possible 
to be described as such locally symmetric K\"ahler manifolds 
under proper settings.
For ${\rm dim}_{\mathbb C} M =2$ case, a star product was explicitly 
given until the second order of derivative, too.
The Grassmann manifolds are typically manifolds of symmetric K\"ahler manifolds.
In Section 5, we also studied the algebraic recursion relations 
for star products for the Grassmann manifolds, too. 
As an example, we constructed a star product for 
${\mathbb C}P^N$, that is equal to the star product 
in \cite{Sako:2012ws}, but it has a different expression. 
\\

{}Before the end of this article, we make two comments.
The first one is about the representation of the noncommutative manifolds given
by the deformation quantization with separation of variables.
In \cite{Sako:2015yrr}, the Fock representation for noncommutative ${\mathbb C}P^N$ is constructed. Using new star products with separation of variables for 
Riemann surfaces given in this article, we can make such kind of Fock representation, similarly. Indeed, the recipe to construct 
the twisted Fock representation is already constructed for general K\"ahler manifolds in \cite{Sako:2016gqb}.
The second comment is about the star products for general Grassmann manifolds.
In this article, star products on ${\mathbb C}P^N$ are given as
an easy case of Grassmann manifolds.
It is expected that explicit star products for the general Grassmann manifolds are obtained by the similar way of this article. 
This problem is left for a future work.\\

\bigskip

\noindent
{\bf Acknowledgments} \\
A.S. was supported in part by JSPS
KAKENHI Grant Number 16K05138.

\vspace{10mm}
\appendix
\section{Proof for Proposition \ref{2dim}}
\label{aa}
$N=2$ and $n=2$ are substituted in Theorem \ref{3_6}.
The results are listed here.
$\vec{\alpha_2},\vec{\beta_2^*}\in \left\{\left(2,0 \right),\left(1,1 \right),\left(0,2 \right) \right\}$
and $i=\left\{ 1,2\right\}$.
\begin{align*}
\lefteqn{\hbar ^2\left(g_{\bar{1}1} \right)^2
=\left(2+\hbar{{R_{\bar{1}}}^{\bar{1}\bar{1}}}_{\bar{1}}\; \right)T^2_{\left(2,0 \right),\left(2,0 \right)}
 +\hbar {{R_{\bar{1}}}^{\bar{2}\bar{2}}}_{\bar{1}}\; T^2_{\left(2,0 \right),\left(0,2 \right)}
+\hbar {{R_{\bar{1}}}^{\overline{2}\bar{1}}}_{\bar{1}} \;T^2_{\left(2,0 \right),\left(1,1 \right)} } \\
 &\hbar ^2\left(g_{\bar{1}2} \right)^2
=\left(2+\hbar{{R_{\bar{1}}}^{\bar{1}\bar{1}}}_{\bar{1}}\; \right)T^2_{\left(0,2 \right),\left(2,0 \right)}
 +\hbar {{R_{\bar{1}}}^{\bar{2}\bar{2}}}_{\bar{1}}\; T^2_{\left(0,2 \right),\left(0,2 \right)}
+\hbar {{R_{\bar{1}}}^{\overline{2}\bar{1}}}_{\bar{1}} \;T^2_{\left(0,2 \right),\left(1,1 \right)} \\
 &2\hbar ^2g_{\bar{1}1} g_{\bar{1}2}
 =\left(2+\hbar{{R_{\bar{1}}}^{\bar{1}\bar{1}}}_{\bar{1}}\; \right)T^2_{\left(1,1 \right),\left(2,0 \right)}
 +\hbar {{R_{\bar{1}}}^{\bar{2}\bar{2}}}_{\bar{1}}\; T^2_{\left(1,1 \right),\left(0,2 \right)}
+\hbar {{R_{\bar{1}}}^{\overline{2}\bar{1}}}_{\bar{1}} \;T^2_{\left(1,1 \right),\left(1,1 \right)}  \\
 &\hbar^2 g_{\bar{1}1}g_{\bar{2}1}
= \left(1+\hbar {{R_{\bar{2}}}^{\overline{2}\bar{1}}}_{\bar{1}}\;  \right)T^2_{\left(2,0 \right),\left(1,1 \right)} 
 +\hbar {{R_{\bar{2}}}^{\bar{1}\bar{1}}}_{\bar{1}}\; T^2_{\left(2,0 \right),\left(2,0 \right)}
+\hbar {{R_{\bar{2}}}^{\bar{2}\bar{2}}}_{\bar{1}}T^2_{\left(2,0 \right),\left(0,2 \right)}  \\
 & \hbar^2 g_{\bar{1}2}g_{\bar{2}2}
=\left(1+\hbar {{R_{\bar{2}}}^{\overline{2}\bar{1}}}_{\bar{1}}\;  \right) T^2_{\left(0,2 \right),\left(1,1 \right)} 
 +\hbar {{R_{\bar{2}}}^{\bar{1}\bar{1}}}_{\bar{1}}\; T^2_{\left(0,2 \right),\left(2,0 \right)}
+\hbar {{R_{\bar{2}}}^{\bar{2}\bar{2}}}_{\bar{1}}T^2_{\left(0,2 \right),\left(0,2 \right)}  \\
 &\hbar^2 g_{\bar{1}1} g_{\bar{2}2}+\hbar^2 g_{\bar{2}1} g_{\bar{1}2}
=\left(1+\hbar {{R_{\bar{2}}}^{\overline{2}\bar{1}}}_{\bar{1}}\;  \right) T^2_{\left(1,1 \right),\left(1,1 \right)} 
 +\hbar {{R_{\bar{2}}}^{\bar{1}\bar{1}}}_{\bar{1}}\; T^2_{\left(1,1 \right),\left(2,0 \right)}
+\hbar {{R_{\bar{2}}}^{\bar{2}\bar{2}}}_{\bar{1}}T^2_{\left(1,1 \right),\left(0,2 \right)}  \\
& \hbar^2 \left(g_{\bar{2}1} \right)^2
=\left(2+\hbar {{R_{\bar{2}}}^{\bar{2}\bar{2}}}_{\bar{2}}\;\right)T^2_{\left(2,0 \right),\left(0,2 \right)}
+\hbar {{R_{\bar{2}}}^{\bar{1}\bar{1}}}_{\bar{2}}\; T^2_{\left(2,0 \right),\left(2,0 \right)} 
+\hbar {{R_{\bar{2}}}^{\overline{2}\bar{1}}}_{\bar{2}}\;T^2_{\left(2,0 \right),\left(1,1 \right)}  \\
& \hbar^2 \left(g_{\bar{2}2} \right)^2
=\left(2+\hbar {{R_{\bar{2}}}^{\bar{2}\bar{2}}}_{\bar{2}}\;\right)T^2_{\left(0,2 \right),\left(0,2 \right)}
+\hbar {{R_{\bar{2}}}^{\bar{1}\bar{1}}}_{\bar{2}}\; T^2_{\left(0,2 \right),\left(2,0 \right)} 
+\hbar {{R_{\bar{2}}}^{\overline{2}\bar{1}}}_{\bar{2}}\;T^2_{\left(0,2 \right),\left(1,1 \right)}  \\
& 2\hbar^2 g_{\bar{2}1} g_{\bar{2}2}
=\left(2+\hbar {{R_{\bar{2}}}^{\bar{2}\bar{2}}}_{\bar{2}}\;\right)T^2_{\left(1,1 \right),\left(0,2 \right)}
+\hbar {{R_{\bar{2}}}^{\bar{1}\bar{1}}}_{\bar{2}}\; T^2_{\left(1,1 \right),\left(2,0 \right)} 
+\hbar {{R_{\bar{2}}}^{\overline{2}\bar{1}}}_{\bar{2}}\;T^2_{\left(1,1 \right),\left(1,1 \right)} \\
 & \hbar^2 g_{\bar{1}1}g_{\bar{2}1}=\left(1+\hbar {{R_{\bar{1}}}^{\overline{2}\bar{1}}}_{\bar{2}} \right)\;T^2_{\left(2,0 \right),\left(1,1 \right)} 
 +\hbar {{R_{\bar{1}}}^{\bar{1}\bar{1}}}_{\bar{2}}\; T^2_{\left(2,0 \right),\left(2,0 \right)} 
+\hbar {{R_{\bar{1}}}^{\bar{2}\bar{2}}}_{\bar{2}}\; T^2_{\left(2,0 \right),\left(0,2 \right)}\\
 & \hbar^2 g_{\bar{2}1}g_{\bar{2}2} =\left(1+\hbar {{R_{\bar{1}}}^{\overline{2}\bar{1}}}_{\bar{2}} \right)\;T^2_{\left(0,2 \right),\left(1,1 \right)} 
 +\hbar {{R_{\bar{1}}}^{\bar{1}\bar{1}}}_{\bar{2}}\; T^2_{\left(0,2 \right),\left(2,0 \right)} 
+\hbar {{R_{\bar{1}}}^{\bar{2}\bar{2}}}_{\bar{2}}\; T^2_{\left(0,2 \right),\left(0,2 \right)}\\
 &   \hbar^2 g_{\bar{2}1}g_{\bar{1}2}+\hbar^2 g_{\bar{1}1}g_{\bar{2}2}
=\left(1+\hbar {{R_{\bar{1}}}^{\overline{2}\bar{1}}}_{\bar{2}} \right)\;T^2_{\left(1,1 \right),\left(1,1 \right)} 
 +\hbar {{R_{\bar{1}}}^{\bar{1}\bar{1}}}_{\bar{2}}\; T^2_{\left(1,1 \right),\left(2,0 \right)} 
+\hbar {{R_{\bar{1}}}^{\bar{2}\bar{2}}}_{\bar{2}}\; T^2_{\left(1,1 \right),\left(0,2 \right)}.
\end{align*}
There are multiple overlapping and tautological equations are omitted.

With (\ref{symr}) these are the same as the following equation.
\begin{align*}
\lefteqn{ \hbar^2\left(
\begin{array}{ccc}
\left(g_{\bar{1}1} \right)^2  & g_{\bar{1}1}g_{\bar{2}1}  &  \left(g_{\bar{2}1} \right)^2    \\
 2g_{\bar{1}1} g_{\bar{1}2} & g_{\bar{2}1}g_{\bar{1}2}+ g_{\bar{1}1}g_{\bar{2}2}& 2 g_{\bar{2}1} g_{\bar{2}2} \\
\left(g_{\bar{1}2} \right)^2  & g_{\bar{2}1}g_{\bar{2}2}  &  \left(g_{\bar{2}2} \right)^2
\end{array}
\right)} \\
 &=\left(
\begin{array}{ccc}
T^2_{\left(2,0\right),\left(2,0 \right)}& T^2_{\left(2,0\right),\left(1,1 \right)}& T^2_{\left(2,0\right),\left(0,2 \right)} \\
T^2_{\left(1,1\right),\left(2,0 \right)}  &T^2_{\left(1,1\right),\left(1,1 \right)}& T^2_{\left(1,1\right),\left(0,2 \right)}\\
T^2_{\left(0,2\right),\left(2,0 \right)} & T^2_{\left(0,2\right),\left(1,1 \right)}& T^2_{\left(0,2\right),\left(0,2 \right)}
\end{array}
\right)
\left(
\begin{array}{ccc}
2+\hbar{{R_{\bar{1}}}^{\bar{1}\bar{1}}}_{\bar{1}}  & \hbar {{R_{\bar{2}}}^{\bar{1}\bar{1}}}_{\bar{1}}  
& \hbar {{R_{\bar{2}}}^{\bar{1}\bar{1}}}_{\bar{2}} \\
\hbar {{R_{\bar{1}}}^{\overline{2}\bar{1}}}_{\bar{1}} &1+\hbar {{R_{\bar{2}}}^{\overline{2}\bar{1}}}_{\bar{1}}  
& \hbar {{R_{\bar{2}}}^{\overline{2}\bar{1}}}_{\bar{2}}\\
 \hbar {{R_{\bar{1}}}^{\bar{2}\bar{2}}}_{\bar{1}} &\hbar {{R_{\bar{2}}}^{\bar{2}\bar{2}}}_{\bar{1}} 
 & 2+\hbar {{R_{\bar{2}}}^{\bar{2}\bar{2}}}_{\bar{2}} 
\end{array}
\right)
\end{align*}
then Proposition \ref{2dim} is proved.



\end{document}